\documentclass[aip,jap
,reprint%
,sd,amsmath,amssymb
,graphicx]{revtex4-1}
\usepackage{graphicx}%
\usepackage{dcolumn}
\usepackage{bm}
\usepackage{amsmath}
\usepackage{pstricks}

\begin{document}
\title{Thermal spin-transfer torque in magnetic tunnel junctions} 
\author{Christian Heiliger}%
 \email{christian.heiliger@physik.uni-giessen.de}
\affiliation{%
I. Physikalisches Institut, Justus Liebig University, Giessen, Germany
}
\author{Michael Czerner}%
\affiliation{%
I. Physikalisches Institut, Justus Liebig University, Giessen, Germany
}
%
\date{\today}
\begin{abstract}
The thermal spin-transfer torque (TSTT) is an effect to switch the magnetic free layer in a magnetic tunnel junction by a temperature gradient only. We present \textit{ab initio} calculations of the TSTT. In particular, we discuss the influence of magnetic layer composition by considering $\text{Fe}_\text{x}\text{Co}_{\text{1-x}}$ alloys. Further, we compare the TSTT to the bias voltage driven STT and discuss the requirements for a possible thermal switching. For example, only for very thin barriers of 3 monolayers MgO a thermal switching is imaginable. However, even for such a thin barrier the TSTT is still too small for switching at the moment and further optimization is needed. In particular, the TSTT strongly depends on the composition of the ferromagentic layer. In our current study it turns out that at the chosen thickness of the ferromagnetic layer pure Fe gives the highest thermal spin-transfer torque.   
\end{abstract}
%
%
\maketitle
%
%
%

\section{Introduction}
\begin{figure}
\includegraphics[width=0.99 \linewidth]{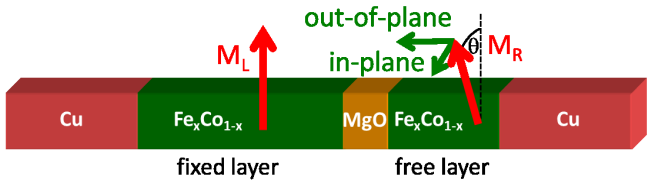}
\caption{
(Color online) Sketch of the MTJ. The barrier (3 monolayers MgO) is sandwiched by two ferromagnetic layers ($\text{Fe}_\text{x}\text{Co}_{\text{1-x}}$) and connected to semi-infinite Cu leads, which are in the bcc-Fe structure. One ferromagnetic layer is the fixed layer (10 monolayers) whereas the other is the free layer (5 monolayers). The relative angle between the two magnetizations $M_L$ and $M_R$ of the two layers is given by $\theta$. The torque acting on the magnetization of the free layer can be decomposed into an in-plane and out-of-plane component. 
}
\label{fig:MTJ}
\end{figure}

In the field of spintronics magnetic tunnel junctions (MTJs) are already used in applications such as read heads in hard disks and storage elements in magnetic random access memories. The tunneling magneto resistance (TMR) effect is exploited in both applications~\cite{moodera95,miyazaki95}. A MTJ consists of a barrier sandwiched between two ferromagnetic layers, a free and fixed layer, shown in Fig.~\ref{fig:MTJ}. Thereby, the size of the current through a MTJ depends on the relative magnetic orientation $\theta$ between the two ferromagnets where basically two stable configuration exists: a parallel ($\theta=0^\circ$) and an anti-parallel ($\theta=180^\circ$) alignment.

Now an effect that can be seen as a counterpart to the TMR is the spin-transfer torque (STT) proposed by Slonczewski~\cite{slonczewski89,slonczewski96} and Berger~\cite{berger96}. Here, a current is driven through the MTJ leading to a switching of the free magnetic layer. This is possible because one magnetic layer acts as a polarizer of the current leading to current with high spin polarization. This current enters the free layer and leads to a precession of the transport electron around the magnetization of the free layer. In turn there is a torque acting on the magnetization of the free layer, which starts to rotate and can finally switch if the current is large enough. A need for a torque is a miss-alignment between the magnetic orientation of the free layer and the polarized current. In reality even thermal fluctuations are enough to guaranty this requirement. Usually, one drives the current through the MTJ by applying a bias voltage. This bias voltage driven STT is widely investigated theoretically \cite{theodonis06,heiliger08a,xiao08,kalitsov09,franz13} as well as experimentally \cite{sankey08,kubota08,wang09}.  

Another way to drive a current through a MTJ is the use of a temperature gradient that leads to a thermo current. This effect of thermal spin-transfer torque (TSTT) is discussed for MgO based MTJs by Jia \textit{et al.} \cite{jia11}. However, this effect has to be distinct from the thermal spin-transfer torque due to magnons proposed by Slonczewski \cite{slonczewski10}. In any case the idea of the TSTT is that the free layer can be switched just by applying a temperature gradient without any bias voltage.

The idea of combining spin-transport with heat is the subject of the new field called spincaloritronics \cite{bauer10,bauer12,silsbee87}. Some of the recently discovered and investigated effects are the spin-Seebeck effect \cite{uchida08,xiao10}, Seebeck spin tunneling \cite{lebreton11}, thermally excited spin currents \cite{tsyplyatyev06}, magneto-Peltier cooling \cite{hatami09}, and magneto-Seebeck effect in metallic multilayers \cite{gravier06}. 
Another effect in MTJs is the tunneling magneto Seebeck effect predicted by us \cite{czerner11} and confirmed experimentally \cite{walter11,liebing11}. This effect can be seen as the analogue to the TMR effect but for the Seebeck coefficient instead of the resistance.

The aim of this paper is to compare the bias voltage driven STT with the thermal STT. In our previous study we showed that the composition of the ferromagnetic layer plays an important role for the TMR effect but has only little influence on the bias voltage driven STT \cite{franz13}. Varying the layer thickness of the ferromagnet leads for both effects to an oscillation that is connected to the Fermi wave length of the ferromagnet \cite{gradhand08,heiliger08a}. However, this cancels if one assumes a thickness fluctuation \cite{heiliger08a}. Now in this paper we discuss the influence of the layer composition and show that there is a strong influence on the TSTT. More general, we discuss in Sec.~\ref{results} that the TSTT is more sensitive to small changes in the tunnel junction than the bias voltage driven STT.

\section{Method}
Our \textit{ab initio} calculations are based on density functional theory using a Green's function based Korringa-Kohn-Rostoker (KKR) method. For the self-consistent calculation of the potentials of the MTJs we use the local density approximation (LDA) as the exchange correlation functional. For the description of the alloys we are using the coherent potential approximation (CPA) \cite{zabloudil05,velicky69}.

For the calculation of the STT we need as an input the exchange field $\vec{\Delta}_i$ of the layer $i$ and the magnetization of the transport electron $\delta \vec{m}_i$ in layer $i$. Then the torque $\vec{\tau}_i$ acting on layer $i$ at a certain energy $E$ is given by \cite{haney07,heiliger08}
\begin{equation}
\vec{\tau}_i(E) = \frac{1}{\hbar} \vec{\Delta}_i \times \delta \vec{m}_i(E) \ .
\label{eq:torque}
\end{equation}  
This means that the torque is a vector perpendicular to the magnetization of the free layer. Consequently, the torque can be decomposed into two components one in-plane and one out-of-plane component (see Fig.~\ref{fig:MTJ}). Thereby, the plane is defined by the magnetizations of the fixed and the free layer. Due to time reversal symmetry the in-plane torque has to vanish at zero bias voltage whereas the out-of-plane torque can be non zero and is related to the interlayer exchange coupling \cite{ralph08}. Whereas the in-plane torque is clearly a torque that leads to switching, the role and importance of the out-of-plane torque is still under discussion. In the following, we focus our investigation on the in-plane component of the torque only.

For the calculation of the torque in Eq.~(\ref{eq:torque}) one needs to determine the magnetization of the transport electron that is the non-equilibrium spin density. Non-equilibrium means in this case that the boundary conditions in the Cu leads are such way that there are only right (left) going electrons in the right (left) Cu leads for electrons originating in the left (right) Cu lead. To do so in the KKR method we use the so called non-equilibrium Green's function (NEGF) or Keldysh method. For further details of this method see Refs. \onlinecite{heiliger08,achilles13}.

For the description of transport properties in alloys, which are described by the CPA, additional so called vertex corrections are necessary. Following Ke \textit{et al.} \cite{ke08} we recently implemented the vertex corrections in our transport formalism and showed the equivalence to a supercell treatment for $\text{Fe}_\text{x}\text{Co}_{\text{1-x}}$ alloys \cite{franz13b}. The advantage of CPA in comparison the a supercell calculation is the fact that arbitrary compositions can be considered. Further, we showed the importance of the vertex corrections for TMR and for the bias voltage driven STT \cite{franz13}.

\section{Results}
\label{results}
\subsection{Comparison of thermal and bias voltage driven spin-transfer torque}
As already mentioned in the following we consider the in-plane component of the torque only. Further, we will discuss the torque acting on the free layer, which means that the layer resolved torque in Eq.~(\ref{eq:torque}) is summed up over all monolayers within the free layer. For the sake of readability we just use $\tau(E)$ for the in-plane component of the torque acting on the free layer.

To obtain the total torque $\tau_{tot}$ acting on the free layer we have to integrate over energy weighted by the occupation function in the lead. To do so we have to distinct between the torque due to electrons originating in the left (L) and in the right lead (R)
\begin{equation}
\tau_{tot} = \int \left ( \tau_{L \rightarrow R} (E) f_L(E,\mu_L,T_L) 
 + \tau_{R \rightarrow L} (E) f_R(E,\mu_R,T_R) \right ) dE  .
\label{eq:torque_total}
\end{equation}
Thereby, $\tau_{L \rightarrow R}(E)$ ($\tau_{R \rightarrow L}(E)$) denote the in-plane torque acting on the free layer for electrons traveling from the left to the right (from the right to the left) lead. The occupation function $f_{L(R)}(E,\mu_{L(R)},T_{L(R)})$ of the left (right) lead depends on the chemical potential $\mu_{L(R)}$ and the temperature $T_{L(R)}$ of the leads.

\begin{figure}
\includegraphics[width=0.99 \linewidth]{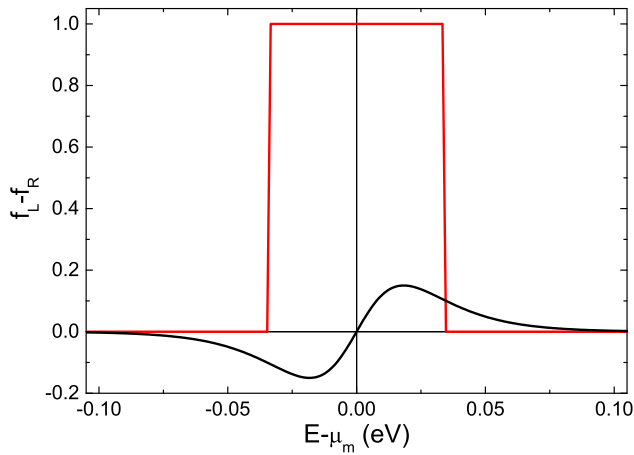}
\caption{
(Color online) Difference of the occupation functions of the right and left lead necessary for Eq.~(\ref{eq:torque_sim}) for two different cases: applying a bias voltage of $68 mV$ (red) and applying a temperature gradient of $T_L-T_R=200 K - 100 K = 100 K$ (black). $\mu_m$ is the averaged electro-chemical potential $\frac{\mu_L+\mu_R}{2}$.
}
\label{fig:occu}
\end{figure}

In ballistic and elastic transport considered here the in-plane torque has to fulfill
\begin{equation}
\tau_{L \rightarrow R}(E) = - \tau_{R \rightarrow L}(E) \ \ ,
\label{eq:torque_equal}
\end{equation} 
which is basically also a consequence from the symmetry argument that the in-plane torque has to vanish with no applied bias voltage and no temperature gradient.
Due to Eq.~(\ref{eq:torque_equal}) Eq.~(\ref{eq:torque_total}) simplifies to
\begin{equation}
\tau_{tot} = \int  \tau_{L \rightarrow R} (E) \left ( f_L(E,\mu_L,T_L) - f_R(E,\mu_R,T_R) \right ) dE \ \ .
\label{eq:torque_sim}
\end{equation}
This equation shows that the difference between the bias voltage driven STT and the thermal STT is just within the occupation functions. Fig.~\ref{fig:occu} shows the difference in the occupation functions when applying a bias voltage and when a temperature gradient is applied. In the first case, the difference leads to a rectangular shape, which is symmetric around the averaged electro-chemical potential. In the case of a temperature gradient the difference of the occupation functions leads to a function that is anti-symmetric with respect to the averaged electro-chemical potential. Usually, the energy dependent in-plane torque has the same sign for each energy as long as the precision is decayed within the ferromagnetic layer. This is fulfilled in our case \cite{heiliger08}.

\begin{figure}
\includegraphics[width=0.99 \linewidth]{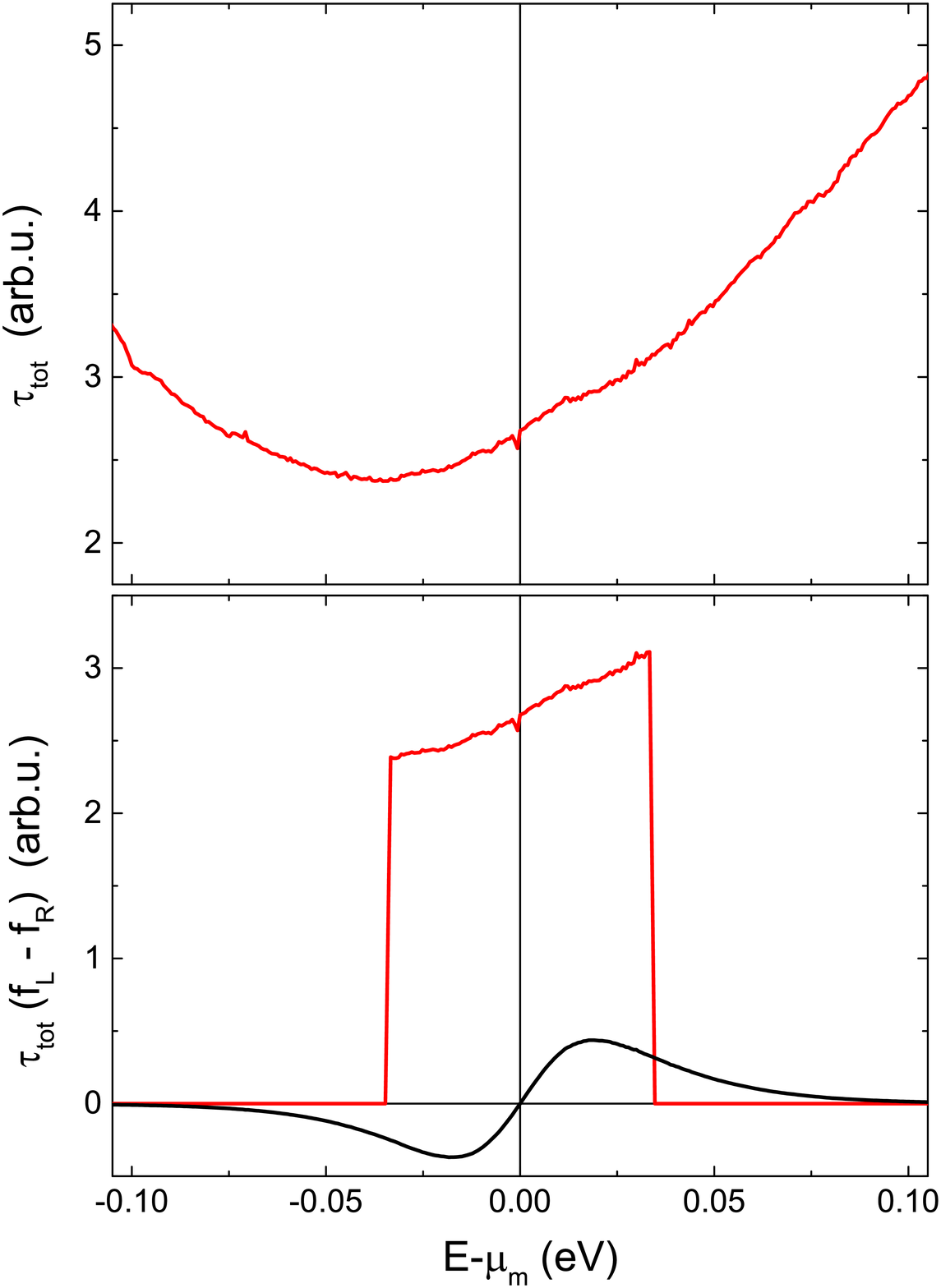}
\caption{
(Color online) Top: Energy dependent in-plane torque $\tau(E)$ acting on the free layer. Bottom: Integrand in Eq.~(\ref{eq:torque_sim}) applying a bias voltage of $68 mV$ (red) and applying a temperature gradient of $T_L-T_R=200 K - 100 K = 100 K$ (black). $\mu_m$ is the averaged electro-chemical potential $\frac{\mu_L+\mu_R}{2}$.
}
\label{fig:integrand}
\end{figure}

Fig.~\ref{fig:occu} already implies that the thermal STT will be much smaller than the bias voltage driven STT in general. 
To illustrate this fact Fig.~\ref{fig:integrand} shows a typical energy dependent in-plane torque, and the integrand in Eq.~(\ref{eq:torque_sim}) for an applied bias voltage and for an applied temperature gradient. Whereas the bias voltage driven STT is basically the area under $\tau(E)$ the thermal STT is basically the asymmetry of $\tau(E)$ with respect to the electro-chemical potential. In particular, for a symmetric $\tau(E)$ the thermal STT will be zero whereas the bias voltage driven STT can be large. This is somehow similar comparing conductance and Seebeck-coefficient. Whereas the conductance is proportional to the area under the transmission function the Seebeck coefficient is related to the asymmetry of the transmission function with respect to the electro-chemical potential~\cite{czerner11}.

This finding already implies that a thermal STT will be possible for only very thin MgO barriers. Jia \textit{et al.}~\cite{jia11} compared the bias voltage driven STT with the thermal STT for different barrier thicknesses and verified this implication on \textit{ab initio} level. In particular, even for a very thin barrier of 3 monolayers the temperature gradient needed for thermal switching is several tens of $K$. \textit{Ab initio} results of the bias voltage driven in-plane torque agree well with experimental results \cite{wang09}. Further, one can estimate the order of magnitude of the torque needed for switching to be about \cite{jia11,wang11}: $4 \cdot 10^{10} Am^{-2} \approx 10 nJ m^{-2}$.

Further, a reduction of the critical switching current can be realized by going to an out-of-plane magnetization of the ferromagnetic layer \cite{ikeda10}. In addition, the free layer should be only a few monolayers thin, because then the torque is already decayed \cite{heiliger08}. Fortunately, going in $\text{Fe}_\text{x}\text{Co}_{\text{1-x}}$ to thin layers an out-of-plane magnetization is achievable \cite{ikeda10}.

In summary, in experiments a MTJ is necessary with a thin MgO barrier, e.g. 3 monolayers and a thin $\text{Fe}_\text{x}\text{Co}_{\text{1-x}}$ layer that shows out-of-plane magnetization. Further, a high temperature gradient has to be realized, e.g. due to a laser pulse. That these requirements can be fulfilled experimentally in general was shown recently by Leutenantsmeyer \textit{et al.}~\cite{leutenantsmeyer13}.

\begin{figure}
\includegraphics[width=0.99 \linewidth]{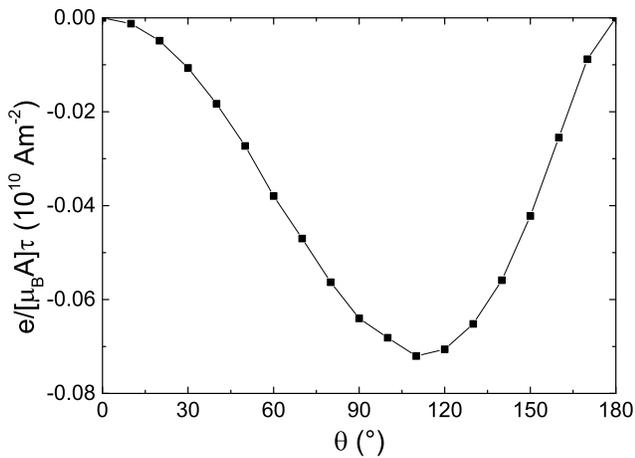}
\caption{
Angular dependence of the thermal STT.
}
\label{fig:angle}
\end{figure}

The very thin MgO barriers can lead also to complicated switching behavior. In particular, the angular dependence of the torque can be different from a typically sine dependence~\cite{jia11}. This is also shown for our structure in Fig.~\ref{fig:angle}. The reason is that for very thin barriers of 3 monolayers MgO  different states contribute to the torque and the symmetry selection of the MgO is not longer fulfilled. Consequently, one has to to analyze the complete angular dependence when changing material parameters. Unfortunately, this is not always possible due to the high computational costs. Nevertheless, it is still useful to analyze the influence of different parameters on the thermal STT at a certain relative angle, which will be done in the next section.

\subsection{Influence of ferromagnetic layer composition}
Recently, we showed that the influence of the composition of the $\text{Fe}_\text{x}\text{Co}_{\text{1-x}}$ layer on the bias voltage driven STT is weak~\cite{franz13}. However, this cannot be automatically expected for the thermal STT. The reason is that in general the thermal STT is more sensitive to changes in the energy dependent torque than the bias voltage driven STT, because the latter is just the area whereas the thermal STT is  a measure for the asymmetry.

\begin{figure}
\includegraphics[width=0.99 \linewidth]{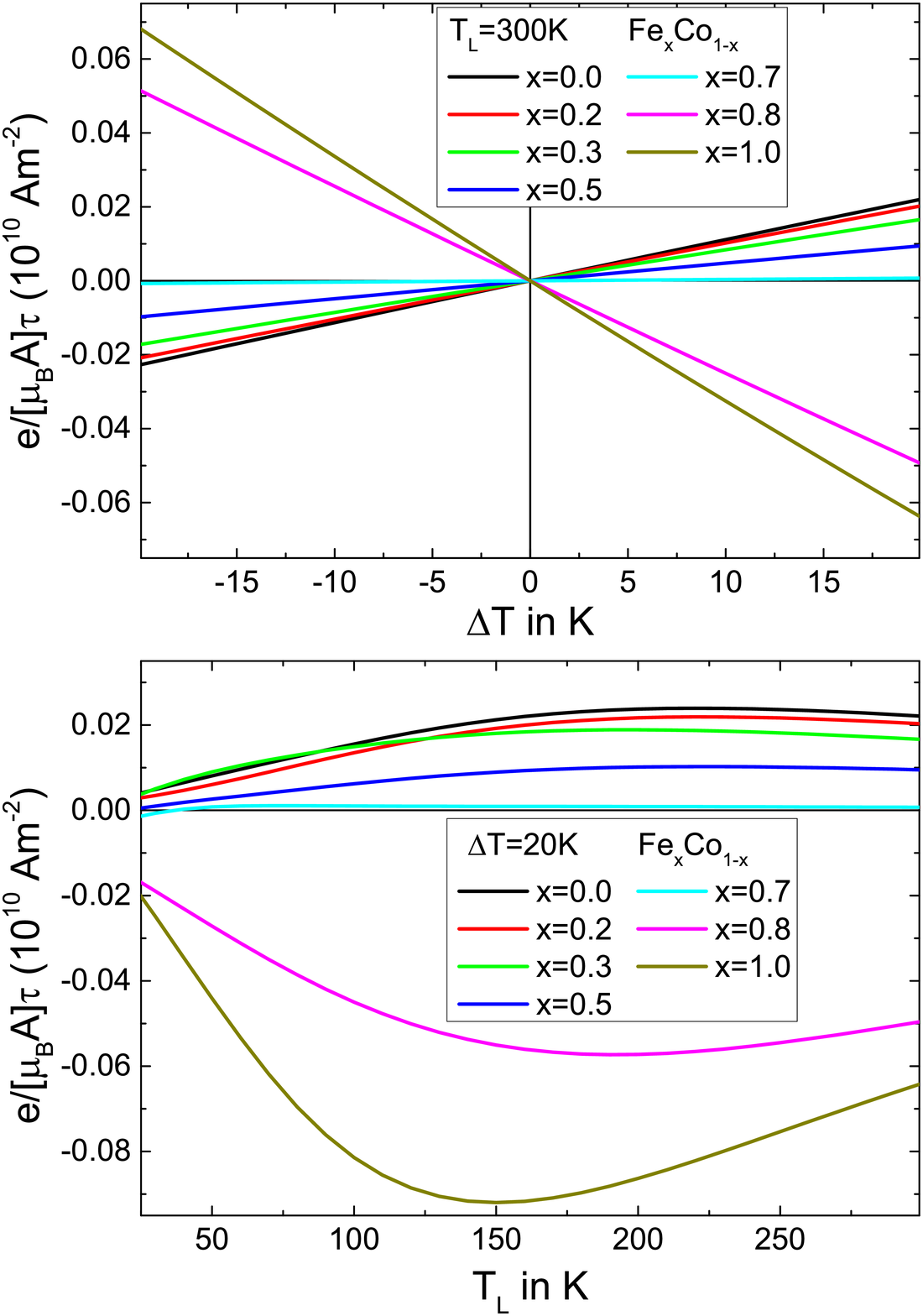}
\caption{
(Color online) Top: Thermal STT at $\theta=90^°$ as a function of the temperature gradient at a fixed temperature of the left lead $T_L=300K$. Bottom: Thermal STT at $\theta=90^°$ and at a temperature gradient of $20K$ as a function of the temperature in the left lead.
}
\label{fig:comp1}
\end{figure}

In the top of Fig.~\ref{fig:comp1} we plot the thermal STT as a function of the temperature gradient for a base temperature of the left lead of 300 K for the different compositions of $\text{Fe}_\text{x}\text{Co}_{\text{1-x}}$. For all compositions we get a linear behavior with increasing temperature gradient. There is a general trend that the torque is largest for pure iron and is decreasing with increasing Co concentration. At around 70\% Fe the thermal STT even vanishes and has than the opposite sign for larger Co concentrations. Now we can fix the temperature gradient and change the base temperature. This is shown in the bottom of Fig.~\ref{fig:comp1}. There is also a general trend that the maximum absolute value of the torque is at lower temperatures. But the change in comparison to room temperatur is quite small and less than 40\%.

\begin{figure}
\includegraphics[width=0.99 \linewidth]{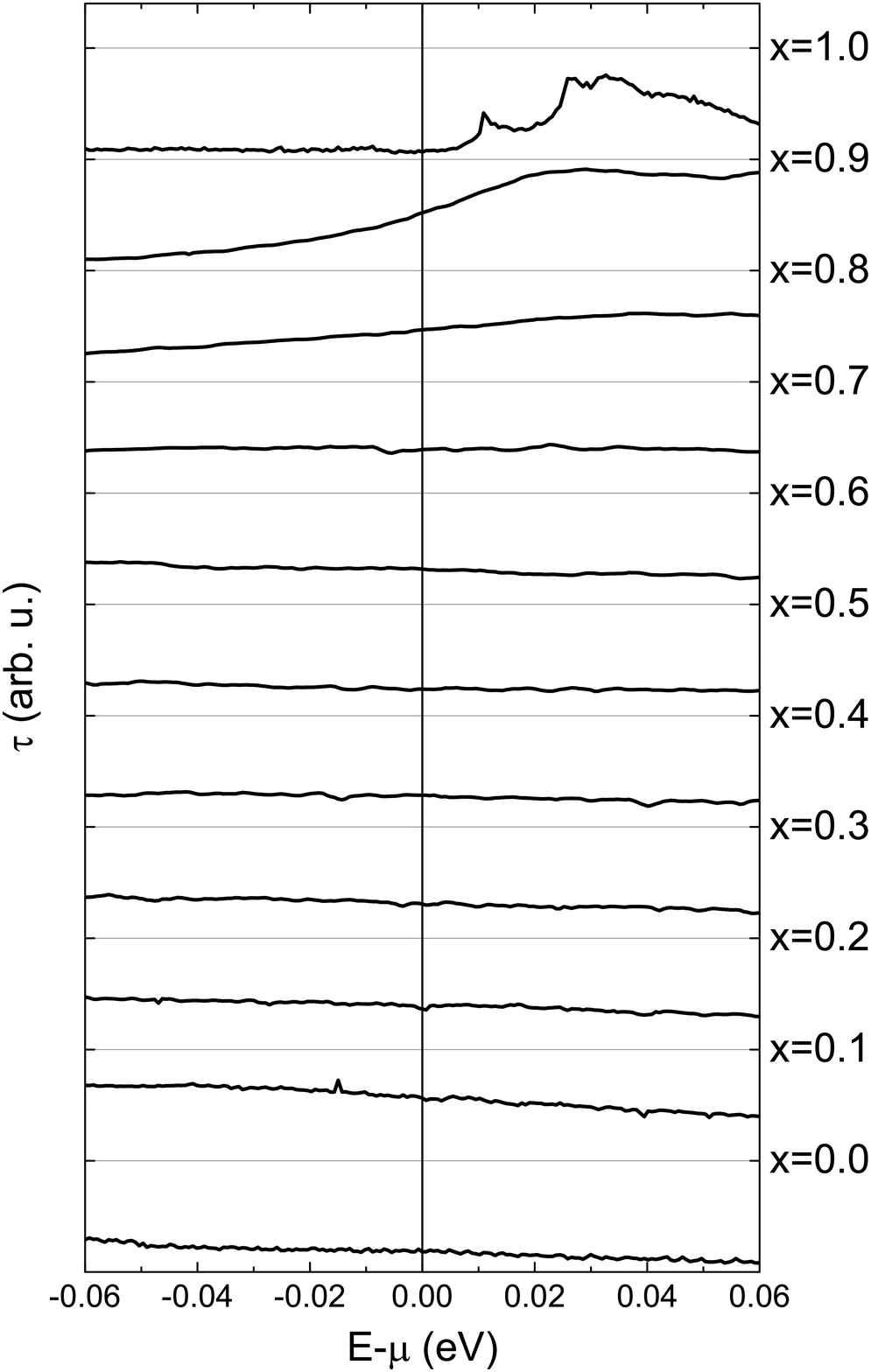}
\caption{
Energy dependent in-plane torque entering in Eq.~(\ref{eq:torque_sim}) for different $\text{Fe}_\text{x}\text{Co}_{\text{1-x}}$ layer compositions.
}
\label{fig:comp2}
\end{figure}

To understand the behavior of the torque with changing composition we analyze the energy dependent torque, which enter in Eq.~(\ref{eq:torque_sim}), in Fig.~\ref{fig:comp2}. For pure Fe there is a pronounced peak above the electro-chemical potential, which leads to strong asymmetry and eventually to a high thermal STT. With increasing Co concentration this peak is shifted to lower energies and is decreasing in amplitude. The shift can be understand simply by the fact that with changing the composition the Fermi level is basically shifted. The argument can be used also to explain the composition dependence of the TMR \cite{franz13} and of the tunneling magneto Seebeck effect \cite{heiliger13}.

However, there are still several issues that are not covered in this manuscript. We fixed the thickness of both ferromagnetic layers. This can lead to quantum well states, which could lead to oscillation in the bias voltage driven STT \cite{heiliger08}. In principle, this could also happen for the thermal STT and could depend on the composition. For the alloys there is additional scattering, which is basically described by the vertex corrections. This can lead to a slightly suppression of these quantum well states. In experiments these quantum well states would be visible if the sample would be perfect. However, due to imperfection, e.g. interface roughness, thickness fluctuations, and defects, these effects will be suppressed. Therefore, we plan in the future to investigate the influence of the layer thickness and of disorder on the thermal STT.

\section{Conclusion}
The thermal STT is a small effect in comparison to the bias voltage driven STT. To achieve a thermal switching it is necessary to use a very thin MgO barrier of not more than 3 monolayers and a very thin free ferromagnetic layer that have out-of-plane magnetization. These requirements are experimentally in principle fulfilled~\cite{leutenantsmeyer13}. However, estimations of the switching temperature based on \textit{ab initio} calculations claim that one need a temperature gradient of up to several ten $K$~\cite{jia11}. This is a large gradient, which is maybe achievable with short laser pulses \cite{leutenantsmeyer13}. Nevertheless, further optimization may be required. Here we show that for pure Fe we expect the largest thermal STT, but in experiment usually compositions of about $\text{Fe}_{\text{0.7}}\text{Co}_{\text{0.3}}$ are used. However, the role of the free layer thickness is still unclear. Further, when we compare our results for pure Fe with the published results by Jia \textit{et al.}~\cite{jia11} our thermal STT is one order of magnitude smaller. Origins maybe the free layer thickness, the atomic positions at the Fe/MgO interface, and the exact position of the Fermi level, in particular, within the MgO gap and at the Fe/MgO interface. In conclusion, there is still a number of unknown influences that need to be investigated. To observe thermal switching an optimization of the thermal STT is needed.

\begin{acknowledgments}
We acknowledge support from DFG SPP 1386 and DFG grant HE 5922/1-1.
\end{acknowledgments}

%
%
%


\end{document}